\documentstyle[prl,floats,aps,epsfig]{revtex}
\input{epsf.sty}
\draft

\begin{document}

\twocolumn[\hsize\textwidth\columnwidth\hsize\csname @twocolumnfalse\endcsname

\title{The 3D Grazing Collision of Two Black Holes}

\author{
Miguel Alcubierre${}^{(1)}$,
Werner Benger${}^{(1,2)}$,
Bernd Br\"ugmann${}^{(1)}$,
Gerd Lanfermann${}^{(1)}$,
Lars Nerger${}^{(1)}$, \\
Edward Seidel${}^{(1,3)}$, and
Ryoji Takahashi${}^{(1)}$
}

\address{
$^{(1)}$ Max-Planck-Institut f{\"u}r Gravitationsphysik,
Am M\"uhlenberg 1, D-14476 Golm, Germany
}
\address{
$^{(2)}$ Konrad-Zuse-Zentrum f\"ur Informationstechnik Berlin, 
Takustrasse 7, D-14195 Berlin, Germany
}
\address{
$^{(3)}$ National Center for Supercomputing Applications,
Beckman Institute, 405 N. Mathews Ave., Urbana, IL 61801
}

\date{\today; AEI-2000-080}

\maketitle

\begin{abstract}
  We present results for two colliding black holes (BHs), with angular
  momentum, spin, and unequal mass. For the first time gravitational
  waveforms are computed for a grazing collision from a full 3D
  numerical evolution. The collision can be followed through the
  merger to form a single BH, and through part of the ringdown period
  of the final BH.  The apparent horizon is tracked and studied, and
  physical parameters, such as the mass of the final BH, are computed.
  The total energy radiated in gravitational waves is shown to be
  consistent with the total mass of the spacetime and the final BH
  mass. The implication of these simulations for gravitational wave
  astronomy is discussed.
\end{abstract}


\narrowtext
\vskip2pc]


The collision of two black holes (BHs) is considered by many
researchers to be a primary candidate for generating detectable
gravitational waves, and hence is the focus of attention for many
research groups worldwide. As the first generation of gravitational
wave detectors \cite{Schutz99}, with enough sensitivity to
potentially detect waves, is coming online for the first time within a
year, the urgency of providing theoretical information needed not only
to interpret, but also to detect the waves, is very high.  However,
even in axisymmetry, the problem has proven to be extremely difficult,
requiring nearly 20 years to solve in even limited cases (e.g.\
\cite{Smarr77,Shapiro92a,Anninos93b,Anninos94b,Matzner95a}).
In full 3D progress has been rather slow
due to many factors, including (but not limited to) unexpected
numerical instabilities, limited computer power, and the difficulties
of dealing with spacetime singularities inside BHs.  
The first true 3D simulation of spinning and moving BHs was
performed in \cite{Bruegmann97}.  In \cite{Bruegmann97}, the two BHs
start out very close to each other, closer than the separation
for the last stable orbit, and the evolution proceeds through parts of
the plunge and ring-down phase of a ``grazing collision'' within a
very short time interval. The spacetime singularities are dealt with
by a particular choice of coordinates, singularity avoiding slicing
and vanishing shift. 

BH excision~\cite{Thornburg87,Seidel92a} has allowed improvements in
the treatment of the spacetime singularities to the extent that highly
accurate simulations of single BHs can be carried
out~\cite{Anninos94e,Anninos94c,Cook97a,Gomez98a,Alcubierre00a} and
recent applications to the grazing collision of BHs show
promise~\cite{Brandt00}. One of the key limiting factors in the
existing two approaches to the grazing collision is the achievable
evolution time for which useful numerical data can be obtained, which
due to numerical problems has been limited to $7M$ in
\cite{Bruegmann97}, and to about $9M$--$15M$ in \cite{Brandt00}. Here
time is measured in units of the total ``ADM'' mass $M$ of the system
as opposed to using the bare mass $m$ of one of the BHs.  Note that
the grazing collision of black holes without spin has also been
studied in the close limit approximation \cite{Nollert97,Khanna99a,Gleiser00}.

In this paper we consider singularity avoiding slicing. We combine the
application of a series of recently developed and tested physics
analysis tools and techniques with significant progress made in
overcoming the problems mentioned above, including the use of much
more stable formulations and much greater computer power. Early,
preliminary results from this series of simulations have been
presented in \cite{Seidel99b,Bruegmann99b}, but we now provide the
first detailed physics analysis not possible previously.


We compute BH initial data of the puncture type~\cite{Brandt97b},
corresponding to two BHs in orbit about each other, with
unequal masses, linear momentum, and individual spins on each BH. 
The construction of such data sets, which involves solving the
non-linear elliptic Hamiltonian constraint equation numerically, is described
in~\cite{Brandt97b}. A detailed survey of a sequence of such data sets
including various physical properties is discussed
in~\cite{Nerger00}. In this paper we choose punctures for each BH on
the $y$-axis at $\pm 1.5m$, masses $m_1 =1.5m$ and $m_2=m$, linear
momenta $P_{1,2} = (\pm 2,0,0) m$, and spins $S_1=(-1/2,0,-1/2)m^2$ and
$S_2=(0,1,-1)m^2$. Note that the linear momentum is perpendicular to
the line connecting the BHs, equal but opposite for a vanishing net
linear momentum, and that the spins are somewhat arbitrarily chosen to
obtain a general configuration.

For this case, an asymptotic estimate for the initial ADM mass is
$M=3.22m$. Solving the Hamiltonian constraint leads to a larger value than the 
Brill-Lindquist mass of $m_1+m_2=2.5m$.
The angular momentum for puncture data is given by (independent of the
solution to the Hamiltonian constraint)
$\vec J = 2 \vec d_1 \times \vec P_1 + \vec S_1 + \vec S_2$, 
where $\vec d_1$ is the vector from the origin to the first puncture.
The total angular momentum is therefore $J = 7.58m^2$, which
corresponds to an angular momentum parameter of $a/M = J/M^2 = 0.73$.
In this configuration the individual spins increase the total angular
momentum, so we call it the ``high-J'' case. The following discussion
refers exclusively to this one data point in parameter space, except
that when discussing waveforms below we compare the high-J case with
data where the individual spins vanish (medium-J, $M=3.00 m$, $J=6.00
m^2$, $a/M=0.67$) or where $S_1
\rightarrow -S_1$ and $S_2\rightarrow -S_2$ (low-J, $M=3.07 m$,
$J=4.64 m^2 $, $a/M=0.49$).


This initial data is evolved with evolution equations of the ``BSSN''
family \cite{Baumgarte99,Shibata95}, using the implementation that we
developed and tested for the collapse of strong gravitational waves to
BHs in \cite{Alcubierre99b}. We discuss some reasons why
certain variable choices and certain combinations of the evolution
equations with the constraints can lead to more stable evolutions than
the traditional ``ADM'' system in
\cite{Alcubierre99e}, and we do observe a significant improvement in
numerical stability in practice. We use radiative boundary conditions
for the outer boundary. The coordinate singularities at the BH
punctures are handled as in \cite{Bruegmann97,Anninos94c} by a time
independent conformal factor. We solve the maximal slicing condition
on the initial slice and then use the so-called 1+log slicing for the
lapse and vanishing shift during the evolutions.


\begin{figure}
\epsfxsize=85mm
\epsfysize=50mm
\epsfbox{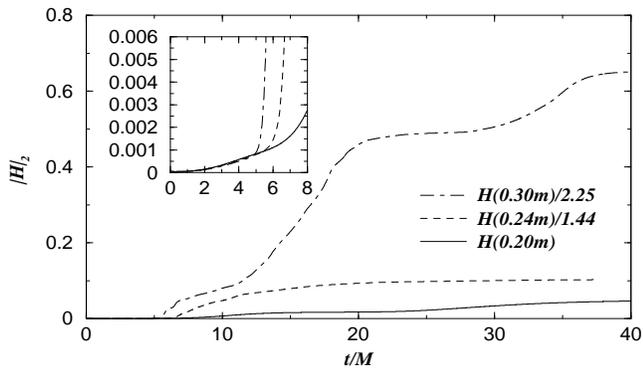}
\caption{Root-mean-square value of the Hamiltonian
constraint on a centered cube with outer boundary at $38m$ and
a gridspacing of $0.30m$, $0.24m$, $0.20m$. The curves are rescaled so that 
they coincide for second order convergence.}
\label{fig:hamil}
\end{figure}

The computer simulations were carried out on a 3D cartesian grid. On a
256 processor SGI/Cray Origin 2000 machine at NCSA we were able to run
simulations of $387^{3}$, which take roughly 100GB of memory (to our
knowledge this makes them the largest production numerical relativity
simulations to date).  A good balance between resolution in the inner
region and distance to the outer grid boundary was achieved for a grid
spacing of $0.2m$, which puts the outer boundary for a centered cube
at about $38m$ or about $12M$.
All said, the combination of resolution, outer boundary location and
treatment, coordinate choice, evolution system and puncture method for
the BHs allows evolution times past $30M$. The lowest quasi-normal
mode of the ring-down phase of the final rapid Kerr BH has a period of
about $13M$ ($17M$ for Schwarzschild), therefore evolution times of
$30M$ or more are a prerequisite for wave extraction, which was not
possible in \cite{Bruegmann97,Brandt00}.  The simulations do not crash
at that time, but as we will discuss now, the numerical data becomes
degraded due to effects of the outer boundary and due to
grid-stretching (i.e.\ large metric gradients) in the vicinity of the
BHs.
Fig.\ \ref{fig:hamil} shows the root-mean-square value of the
Hamiltonian constraint over the entire grid for different resolutions
but same outer boundary location. The inset shows clean global second
order convergence up to about $6M$. A local analysis shows that there
are large contributions to this average from inside the horizon, and
that smaller errors intrude from the outer boundary, but the code is
convergent beyond $30M$.


\begin{figure}
\epsfxsize=85mm
\epsfbox{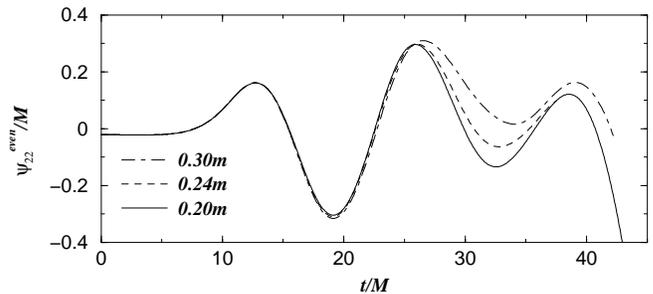}
\caption{Waveform at resolutions $0.30m$, $0.24m$, $0.20m$.}
\label{fig:Qconv}
\end{figure}

\begin{figure}
\epsfxsize=85mm
\epsfysize=45mm
\epsfbox{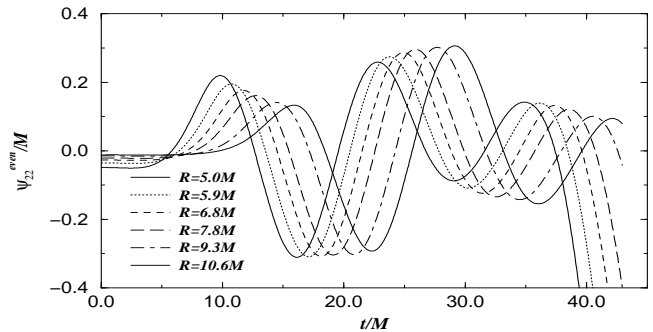}
\caption{Mode $l=m=2$ of the even Zerilli function extracted for
different radii as a function of time.
A wave that develops after the BHs collide is propagating out. 
}
\label{fig:QmanyR}
\end{figure}

\begin{figure}
\epsfxsize=85mm
\epsfysize=50mm
\epsfbox{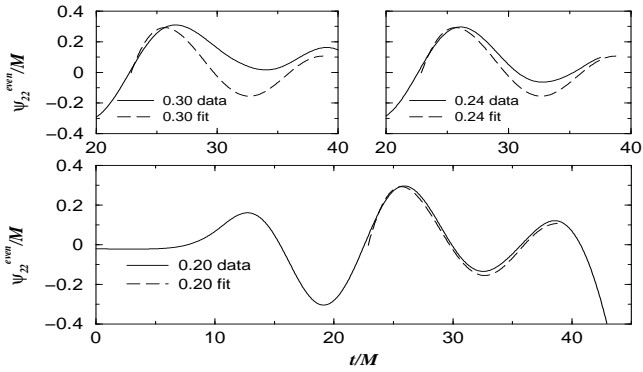}
\caption{A fit to the quasi normal mode determined by $M$ and $a$ shows 
good agreement in the frequency and decay rate at late times for a
resolution of $0.2m$.}
\label{fig:wavefit}
\end{figure}

Since a main result of the simulations are waveforms, the most
relevant measure and often most stringent criterion for numerical
quality is convergence in the waveforms. We use the gauge
invariant waveform extraction technique, developed originally by
Abrahams~\cite{Abrahams88} and applied to the 3D case
in~\cite{Allen98a}, to extract gravitational wave modes of arbitrary
${\ell,m}$.  As shown in~\cite{Allen98a,Baker99a}, this technique can
be used on numerically evolved 3D distorted BH spacetimes to produce
very accurate waveforms away from the BH, even if errors are rather
large near the horizon.  Here, we extract for example the
nonaxisymmetric $\ell=m=2$ mode, expected to be one of the most
important modes in binary BH coalescence~\cite{Flanagan97a}.
Fig.\ \ref{fig:Qconv} shows for three resolutions the Zerilli function
$\psi^{even}_{22}(t)$ extracted at $R = 7.8M$.  
Up to a time $t \approx 30M$ the dependence on resolution is rather
small, which suggests that the resolution reaches the convergent regime.

\begin{figure}
\epsfxsize=87mm
\epsfysize=60mm
\epsfbox{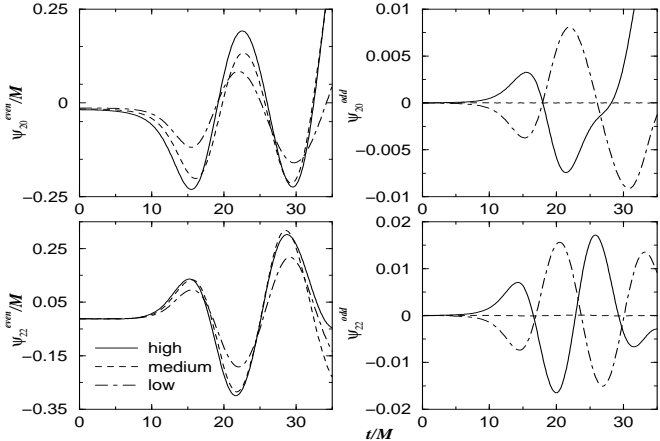}
\caption{Even and odd wave parts showing differences depending on high,
medium, and low-J data.}
\label{fig:Qhimedlo}
\end{figure}

In Fig.~\ref{fig:QmanyR}, we show a sequence of extracted waves at
different radii, obtained by integration over the corresponding
coordinate spheres, as a function of time. The outermost detectors
show late time problems due to spurious signals propagating in from
the outer boundary, while the inner detectors are affected by the
closeness to the strong field region.  Note that these methods assume
a Schwarzschild background, but they can be applied on a rotating BH,
the primary effect being an offset depending on the rotation parameter
$a$ \cite{Brandt94c}.  In Fig.~\ref{fig:wavefit}, we show the $l=m=2$
even parity wave for the detector at $R = 7.8M$ and a match to the
corresponding lowest quasi-normal mode plus the first overtone. The
values for $M$ and $a$ determine the quasi-normal frequency, while the
amplitude and the offset in time are fitted. The observed period is
$13M$, which is consistent with a final distorted Kerr BH with $a/M
= 0.73$. Gravitational waves carry away energy and momentum from the
BHs. For the energy, we have $dE/dt=1/(32\pi) \sum^{\infty}_{l=2}
\sum^{l}_{m=-l} ((d\psi^{even}_{lm}/dt)^2+(\psi^{odd}_{lm})^2).$
Integrating the $l = 2,3,4$ modes up to $t = 35M$, we find $\Delta E =
0.0323m \approx 1\% M$.

One of the potential insights from the detection of gravitational
waves is the determination of the orientation of spins in relation to
the orbital motion. Fig.\ \ref{fig:Qhimedlo} shows the wave signature
for the high, medium, and low-J data. For vanishing spins (medium-J),
we note that $\psi^{odd}_{20}$ is zero within numerical accuracy,
while $\psi^{odd}_{22}$ shows an oscillation with amplitude
$3\times10^{-5}$. A comparison with the methods of \cite{Gleiser00}
would be useful.


\begin{figure}
\noindent
\epsfig{file=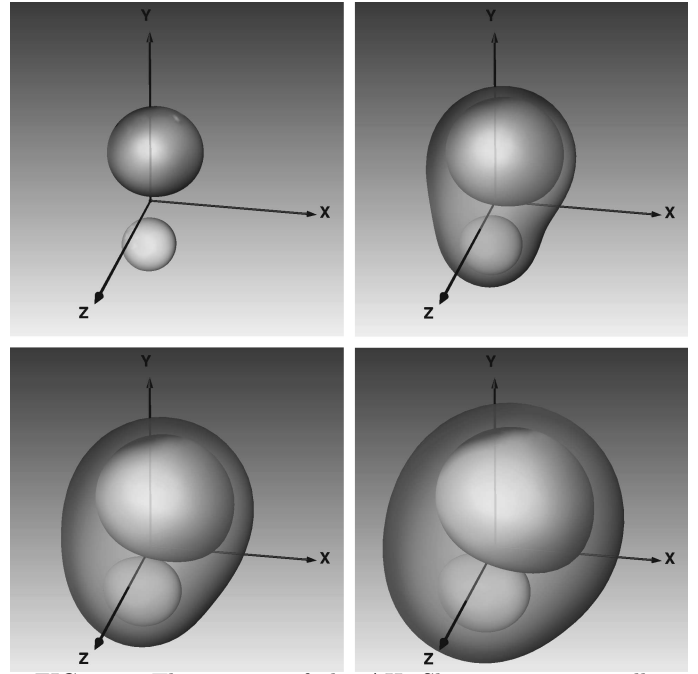,width=90mm}
\caption{
The merger of the AH. Shown are marginally trapped surfaces at times 
$2.5M$, $3.7M$, $5.0M$, and $6.2M$. The apparent horizon is
the outermost of these surfaces.
}
\label{fig:ah3d}
\end{figure}

While it will be the waves that we can observe on earth directly, it
is also interesting to compute the apparent horizon in the grazing BH
collision. During the evolution we use a 3D apparent horizon (AH)
finder described in~\cite{Alcubierre98b} to track the location of the
horizon. In principle, the event horizon can also be located by
techniques developed in Ref.~\cite{Anninos94f}, but we do not yet know
whether a single event horizon is present on the initial slice in this
data set.  Fig.~\ref{fig:ah3d} shows the AH during a grazing
collision.

\begin{figure}
\epsfxsize=82mm
\epsfysize=40mm
\epsfbox{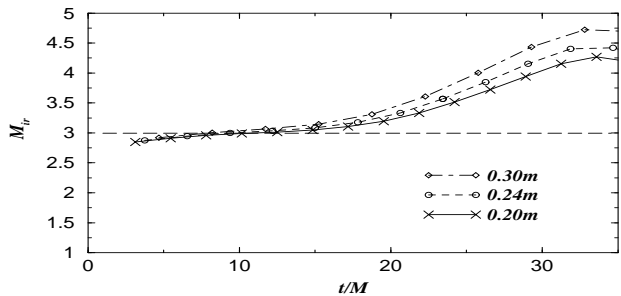}
\caption{
We show the evolution of $M_{ir}$ for the high-J configuration. As the
BHs merge, the area grows, and then begins to level, as the final BH
goes into the ringdown. But numerical error associated
with the grid stretching effects causes a spurious growth in the area,
familiar in previous 2D and 3D studies.  However, one can estimate the
mass of the final BH, as shown by the dashed line.}
\label{fig:ahmass}
\end{figure}

We compute the BH mass $M_{AH}$ and compare with the ADM mass of the
initial data and the radiated energy to assess the overall energy
accounting.  In Fig.~\ref{fig:ahmass}, we show the result of the
calculation of the so-called irreducible mass as a function of time,
defined as \mbox{$M_{ir} = \sqrt{Area_{AH}/16\pi}$}.  The horizon mass
$M_{AH}$ can be determined through the formula \mbox{$M_{AH}^{2} =
  (M_{ir})^{2} + J^{2}/(2M_{ir})^{2}$}, where we use $J = 7.58m^2$ of
the initial data.  The observed upward drift in $M_{ir}$ may be
curable by excision or better coordinate conditions, but even in the
present case we can estimate the final mass of the BH to be $M_{ir}
\approx 3.0m$ and $M_{AH}\approx 3.3m$ in this simulation.  Comparing
this to the initial ADM mass of the spacetime, $M=3.22m$, we find
consistency in the overall energy accounting from independent physical
measurements.  The fraction of the total energy, $1\%$, that is
carried away in the gravitational waves falls within the error
estimate of this energy balance.


In conclusion, these results indicate that for the first time we are
indeed able to simulate the late merger stages of two BHs colliding,
with rather general spin, mass, and momenta, and that we can begin to
examine the fine details of the physics.  Studies of apparent
horizons, waveforms, and asymptotic properties show consistency in the
analysis across strong field, near zone, and far field regions.
Without more advanced techniques, such as BH excision, these
simulations will be limited to the final merger phase of BH
coalescence.  But while that is under development, we can take
advantage of our capabilities and explore this phase of the inspiral
now.  Our goal is several fold: (a) to explore new BH physics of the
``final plunge'' phase of the binary BH merger, (b) to try to
determine some useful information relevant for gravitational wave
astronomy, and (c) to provide a strong foundation of knowledge for
this process that will be useful when more advanced techniques are
fully developed. A new development is the Lazarus project, which
provides an interface between full numerical relativity simulations
and perturbative evolutions \cite{Baker2000b}, and an interface to the
post-Newtonian inspiral phase is planned. At this time Lazarus allows
us to move a finite time interval of full numerical evolution into the
early stages of the merger, and to perform the ring-down calculation
efficiently once the final BH enters the perturbative regime.  
When these techniques and excision are used to extend the ability of
the community to handle the collision of two BHs starting from the
late orbital phase, it will be important to have an understanding of
details of the most violent merger phase in advance, both as a testbed
to ensure that results are correct, and because the understanding we
gain may be useful in devising the appropriate techniques for longer
term evolution.

\noindent {\bf Acknowledgments.} 
We thank in particular G.~Allen, J.~Baker, M.~Campanelli, T.~Goodale,
C.~Lousto, and many colleagues at the AEI, Washington University,
Universitat de les Illes Balears, and NCSA for the co-development of
the Cactus code, and for important discussions and assistance.
Calculations were performed at AEI, NCSA, SDSC, RZG, and
ZIB.

\bibliographystyle{prsty}
\bibliography{bibtex/references}

\end{document}